\documentclass[11pt]{article}
\usepackage{moriond,epsfig}
\usepackage{xspace}

\def\be{\begin{equation}}
\def\ee{\end{equation}}
\def\bea{\begin{eqnarray}}
\def\eea{\end{eqnarray}}

\newcommand{\cdf}{CDF\xspace}
\newcommand{\dzero}{D\ensuremath{0\hspace{-0.075in}/}\xspace}
\newcommand{\ttbar}{\ensuremath{t\bar{t}}\xspace}
\newcommand{\sigmattbar}{\ensuremath{\sigma_{t\bar{t}}}\xspace}

\newcommand{\invfb}{\ensuremath{\mbox{fb}^{-1}}\xspace}
\newcommand{\tZq}{\ensuremath{t\rightarrow Zq}\xspace}
\newcommand{\chisq}{\ensuremath{\chi^{2}}\xspace}
\newcommand{\Ht}{\ensuremath{H_{T}}\xspace}
\newcommand{\cts}{\ensuremath{cos(\theta^{*})}\xspace}
\newcommand{\unit}[1]{\ensuremath{\mathrm{\:#1}}}

\begin{document}
\vspace*{4cm}
\title{MEASUREMENTS OF TOP QUARK PROPERTIES AT THE TEVATRON}

\author{R. Eusebi \\ 
  (on behalf of the \cdf and \dzero collaborations)
}

\address{Fermi National Accelerator Laboratory,\\
         Batavia 60510-5011, Illinois, USA.
}

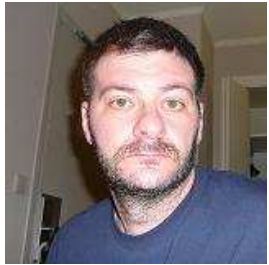
\maketitle
\abstracts{The precise measurement of the top quark properties is a stringent test of the Standard Model of Particles and Fields. This reports presents the latest results from the \cdf and \dzero collaborations with an integrated data sample of up to 2.3\unit{\invfb} }

%----------------------------     Introduction      ---------------------------
%------------------------------------------------------------------------------
\section{Introduction}

The top quark was discovered in 1995 \cite{topCDF,topDzero} by the \cdf and \dzero collaborations. Its large mass suggest it is strongly associated with the mechanism of electro-weak symmetry breaking, and makes it the fermion with the largest coupling to the standard model (SM)-expected, but not yet found, Higgs boson. These reasons make the top quark potentially sensitive to new physics, which  can be revealed through precision measurements of its production and decay properties. 
This letter reports the results of measurements of top quark properties with up to 2 \invfb of data. In general most of the analyses presented here were performed by both, the \cdf and \dzero collaborations, however a single analysis, the most accurate one of either collaboration, would be presented.

%---------------------------- Production Properties ---------------------------
%------------------------------------------------------------------------------
\section{Measurements of Top Quark Properties}
\subsection{Top Charge Asymmetry}

The measurement  of the \ttbar charge asymmetry quantifies the forward-backward asymmetry on the top production. While the theoretical prediction for the magnitude and structure of the asymmetry is effectively unknown, the asymmetry is expected to be low, making the measurement of this asymmetry a sensitive probe for new physics.

This measurement was performed by the \dzero collaboration with a data sample of 0.9\unit{\invfb} using the lepton plus jets \ttbar decay channel. Events are fully reconstructed using the kinematic fitter, which fits the final states jets and leptons to the \ttbar decay hypothesis. From this reconstruction the rapidities of the top ($y_t$) and anti-top ($y_{\bar{t}}$) are obtained.  

The top production asymmetry is defined as $A_{fb} = \frac{N^{f}-N^{b}}{N^{f}+N^{b}}$, where $N^{f}$ and $N^{b}$ represent respectively the number of events in which the signed rapidity of the top is larger and smaller than that of the anti-top.
% The top production asymmetry is defined as $A_{fb} = \frac{N^{\Delta y>0}-N^{\Delta y <0}}{N^{\Delta y>0}+N^{\Delta y <0}}$, where $\Delta y=y_t-y_{\bar{t}}$ and $N^{\Delta y>0}$ ($N^{\Delta y<0}$) represent the number of events in which the signed rapidity of the top is larger (smaller) than that of the anti-top. 
The asymmetry observed in data can be predicted for any model, while taking detector effects into account, by:
\begin{equation}
A_{fb}^{pred} = \int_0^\infty A_{fb}(|\Delta y|) D(|\Delta y|) f(|\Delta y|) d|\Delta y|,
\label{eq:murnf2}
\end{equation}
where $D$ is the dilution due to detector effects, and $f$ is the probability density predicted by the model.  The predicted asymmetry at reconstruction level however, depends strongly on the experimental acceptance, and the event selection criteria was kept as simple as possible.
This analysis~\cite{ChargeAsymm_dzero} provides a dilution function and report limits for lepton plus jets events with exactly four jets, and for lepton plus jets events with five or more jets. In addition, based on the large asymmetry predicted when a proposed lepto-phobic particle $Z'$ decays to \ttbar, this analysis set limits on the $Z'$ production as a function of the mass of the $Z'$.

%---------------------------- Intrinsic Properties ---------------------------
%------------------------------------------------------------------------------
\subsection{Top Quark Charge}

One of the basic quantities that characterize the top quark is its electric charge, which in the SM is expected to have a value of $\frac{2}{3}e$. While a direct measurement of the top charge is not feasible due to its fast decay, the total charge of the decay products can be measured. In the assumption that the top quark decays to a W boson and a b quark, and given the well measured W and b charges, two possibilities arise; the top quark decays to a $W^{+}$ and a b quark, hence having a charge of  $\frac{2}{3}e$, or decays to a $W^{-}$ and a b quark, hence having a charge of $\frac{4}{3}e$. Top quarks with fractional charge of  $\frac{4}{3}e$ have  been proposed in the literature\cite{top_charge} as part of a fourth generation of quarks and leptons.
 
Here we present the CDF result using a 1.5\unit{\invfb} of data in the dilepton and lepton plus jets channel.  The measurement identifies the charge of the two W's and two b-quark's in each data event, and then determines which W and b-quarks decayed from the same parent top quark. The charge of the top is then obtained by multiplying the charge of the W with the charge of the jet associated with a b-quark, obtaining two (W,jet) pairs.  Pairs whose charge product is negative are considered SM-like pairs (SM-like), while those whose product is positive are considered exotic model-like (XM-like) pairs. Based on the total number of SM-like and XM-like pairs, limits can be set on the validity of the two models.

The charge of one W is obtained by identifying the charge of the lepton and the charge of all the b-tagged jets is obtained from the Jet-charge algorithm. A profile likelihood method is used to build a likelihood curve as a function of SM-like events. The probability to incorrectly reject the SM is set, a priori, to 1\%. From Monte Carlo studies the probability of rejecting the SM when the XM is true is found to be 87\%. With 1.5\unit{\invfb} of data CDF observes 124 SM-like pairs, and 101 XM-like pairs, obtaining a p-value of 0.31\%. Since this value is greater than the a priori-chosen value of 1\% the XM model is excluded at 87\% C.L \cite{top_charge_cdf}.

%---------------------------- Decay Properties --------------------------------
%------------------------------------------------------------------------------

\subsection{The ratio of branching ratios $BR(t\rightarrow W^{+}b)/BR(t\rightarrow W^{+}q)$}

Within the SM the top quark decays  to a $W$ boson and a down-type $q$ quark with a rate proportional to $|V_{tq}|^{2}$.  The ratio of the branching ratio top decay to $Wb$ to that to $Wq$ is related with the elements of the CKM matrix by 
\begin{equation}
R=\frac{BR(t\rightarrow W^{+}b)}{BR(t\rightarrow W^{+}q)}=\frac{|V_{tb}|^{2}}{|V_{td}|^{2}+|V_{ts}|^{2}+|V_{tb}|^{2}}
\end{equation}

The average number of  b quarks from the decay of a generic \ttbar decay event directly depends on the value of $R$, hence so does the probability of tagging a jet as coming from a b-quark. The \dzero collaboration measured $R$ simultaneously with the \ttbar cross section, \sigmattbar, based on the distribution of events with 0, 1 and 2 or more b-tags jets using 0.9\unit{\invfb}.A likelihood fit to these two variables is performed simultaneously obtaining $R=0.97^{+0.09}_{-0.08}$ and $\sigmattbar=8.18^{+0.9}_{-0.84}(stat+syst)\pm0.5(lumi)$. The observed value of $R$ is translated to a lower 95\unit{\%} confidence limit by using the Feldman-Cousin ordering principle, obtaining $R>0.79$ at 95\unit{\%}C.L. This value is the best direct limit on $R$ to date. In addition,  assuming $R=|V_{tb}|^{2}$ we obtain $|V_{tb}|>0.89$ at 95\unit{\%}C.L.

%------------------------------------------------------------------------------
\subsection{Flavor Changing Neutral Currents}

In the SM flavor changing neutral currents (FCNC) are allowed at orders higher than tree level. The decay \tZq is very rare, with a branching ratio B(\tZq) of about $10^{-14}$ in the SM, but with the potential to reach values as high as $10^{-2}$ in exotic scenarios involving new physics\cite{XM_TZQ}.

The \cdf collaboration has performed a search for the flavor changing neutral current decay of the top quark \tZq using a data sample corresponding to an integrated luminosity of  1.9\unit{\invfb}. Candidate events are selected by requiring two opposite sign leptons ($e$'s or $\mu$'s), 4 or more jets and a series of optimized cuts.  Events in this signal region are further classified according to whether or not they have a secondary vertex (b-tag). A third sample is used as control and made from rejected events that failed to pass at least one of the optimized requirements.

 The signal is discriminated from the background by exploring kinematic constraints present in FCNC  events. A mass \chisq variable quantifies the consistency of each event with originating from a top quark FCNC decay.  Templates of this variable are generated for the main backgrounds, and the FCNC signal. Shape systematic uncertainties are included in the templates.
The \chisq template fit is implemented as a simultaneous fit to two signal regions and the control region. Assuming  a top quark mass of $175$\unit{GeV/C^2} the expect sensitivity of the measurement is to set an upper limit on $B(\tZq)$ of $5.0\%$. The results of the fit are consistent with the \chisq distribution of the background. An upper limit of $B(\tZq)< 3.7\%$ at 95\% C.L. is obtained using the Feldman-Cousins prescription.

%------------------------------------------------------------------------------
\subsection{$W$ Helicity Polarization from Top Decays}

In the SM the top quark decays via the V-A interaction, almost always to a $W$ and $b$ quark. A different Lorentz structure of the $t\rightarrow W b$ interaction can alter the fractions of $W$ bosons produced in each polarization state from the SM values of $f_{0}=0.69\pm0.01$ and $f_{+}=3.6 10^{-4}$ for the longitudinal and right-handed fraction respectively. The left-handed fraction is assumed to be  $f_{-}=1-f_{+}-f_{0}$.  The polarization of the $W$ can be described using the angle $\theta^{*}$ between the $W$ momentum in the top quark rest frame and the down type fermion momentum in the $W$ rest frame.

The \dzero collaboration has measured the longitudinal and right-handed fractions of the $W$ boson helicity using 1\unit{\invfb} of data.
The \ttbar candidate events are selected according to the dilepton and lepton plus jets topologies. 
Lepton plus jets events are fully reconstructed using the kinematic fitter. Templates of \cts are made for \ttbar with different $W$ polarizations and for the backgrounds, distinguishing between the leptonic and hadronic $W$'s in the event. In hadronically decaying $W$ the down-type quark is randomly assumed to be one of the jets associated with the boson.
Dilepton events have four-fold ambiguity in the reconstruction. The \cts is determined for each of the four combinations and templates made for the \ttbar signal with different $W$ polarizations and for the backgrounds.
A fit is made simultaneously to the three  set of templates  measuring $f_{0}=0.425\pm0.166(stat)\pm0.102(syst)$ and  $f_{+}=0.119\pm0.090(stat)\pm0.053(syst)$. These are the most sensitive measurements of the $W$ polarization to date.

%------------------------------------------------------------------------------
\subsection{Search for a fourth generation $t'$}

Fourth generation $t'$'s are predicted in some SUSY models\cite{XM_TPRIME}, and there is room in the electroweak data to accommodate a heavy Higgs (~500 GeV) without any other new particles. The \cdf collaboration has searched for a heavy top ($t'$) quark pair production decaying to Wq final states in 2.3\invfb in the lepton plus jets data sample without b-tagging requirements.  The $t'$ is assumed to be produced in pairs via the strong interaction, to have mass greater than the top quark, and to decays promptly and only to $Wq$ final states.

Two variables are directly related to the mass of $t'$; the total transverse energy of the event (\Ht), and the reconstructed mass of the $t'$ ($M_{reco}$) as obtained from the  kinematic fitter. To discriminate the new physics signal from standard model backgrounds a set of 2D-templates of the main backgrounds, as well as different mass $t'$'s, are constructed in the (\Ht,$M_{reco}$) plane. For a given $t'$ mass, the observed data is fitted to the background 2D-template and to the 2D-template of the given $t'$ mass, to set limits on the $t'$ production. Using a specific $t'$ model\cite{XM_TPRIME2}, this analysis exclude $t'$ with masses below 284\unit{GeV/C^2} at 95\% C.L.

\section*{Acknowledgments}
I would like to thank the \cdf and \dzero collaborations for the large amount of work, in particular  the authors of all the analyses shown here for their critical input, and the respective top group conveners for their support and advice. I also thank the organizers of the Moriond QCD 2008 conference.

\end{document}